# Multi-round Dynamic Group Decision Making Method On 2-Dimension Uncertain Linguistic Variables

Yukun Zhang

The Chinese University Of HongKong    215010026@link.cuhk.edu.cn

**Abstract.** The language evaluation information of dynamic multi-round group decision-making method at present is based on the one-dimension language variable. The 2-Dimensional Uncertain Linguistic Generalized Weighted Aggregation (DULGWA) are powerful tools to express the fuzzy or uncertain information. But the multi-attribute group decision based on one-dimensional language evaluation information adopts single-stage and static evaluation methods. In this paper, we combine dynamic interactive group decision making methods with two-dimensional language evaluation information, proposing multi-round dynamic group decision making method based on two-dimensional language evaluation information. This method not only introduces multi-attribute decision-making method into the dynamic field, but also overcome the limitation of the evaluation information as one-dimensional language, improving greatly the universal applicability of interactive evaluation methods. Then we use vector fitting method to synthesize expert evaluation information of multiple rounds and finally find the preference of decision makers. One example is given to verify the developed approach and to demonstrate its effectiveness.

# 1.Introduction

Group decision-making methods have wide applications in many fields, such as economic, science, culture and other aspects .As for group decision-making problems ,decision maker have difficulties expressing the attribute values by the real numbers owing to the complexity and uncertainty of human thinking .while it may be easier to express the attribute values by linguistic information. These problems are called Uncertain Multiple Attribute Decision Making (UMADM) problems. The research of Multiple Attribute Decision Making(MADM) problems theory have made great progress up to now.As for UMADM theory, Zedeh [1] firstly proposed the definition of the linguistic variables. Zhu et al.[3] introduced the 2-dimension linguistic evaluation information to solve the fuzzy problems .which means I class linguistic information and II class linguistic information are used to evaluate the objects. I class linguistic information describe the evaluation results of attributes given by
 decision makers,while II class linguistic information gives the subjective evaluation of decision makers on the reliability of there evaluation,hence we call the 2-dimension linguistic assessment information as 2-dimension linguistic variables, which are more accurate to express the opinion of decision makers for attribute values.

Up to now,there are some studies of 2-dimension linguistic information method, Zhu et al.[3] introduced the definition of 2-dimension linguistic information and apply the framework into subjective evidencial reasoning in the area of group decision making, but this method does not deal with the multi-criteria decision problems. Liu and Zhang [4] further extend 2-dimension linguistic variables to 2-dimension uncertain linguistic variables in which the attribute weight is unknown and 2-dimension uncertain linguistic information is uncertain, and propose a new method to deal with multiple attribute group decision making problems. Liu and Yu [5]   proposed 2-dimension uncertain linguistic power generalized aggregation operators(2DULPGA) and 2-dimension uncertain linguistic power generalized weighted aggregation operators(2DULPGWA) combining the power average(PA) operator with 2-dimension linguistic information, and discuss some properties and special cases of them. Then they proposed group decision making method based on some power generalized aggregation operators with 2-dimension uncertain linguistic information.

There is no doubt that the research on multiple attribute of group decision making problems based on 2-dimension linguistic information have great significance, but in the real decision making problems,it does not consider the dynamic and interactive case. At present, the main research achievements based on dynamic group decision making method have ,made the fruitful achievements. For example,Zhang [5] discusses a method of group evaluation based on multi-phase dynamic interactive and then defines stability and consistency of indicator indexes to explore the interaction.

Obviously,there are two shortcomings of the group decision-making method.Firstly, the main research of traditional dynamic group decision making method is focus on the one dimension and the number value and it is difficult for one dimension information to express the fuzzy assessment . And the multi-attribute decision making on the 2-dimension uncertain linguistic information just at present stop at the single stage and do not take the dynamic decision-making perspective into consideration. In view of the above two shortcomings ,we combine the 2-dimension linguistic variables with dynamic group evaluation and propose the multi-round dynamic group decision making method on 2-dismension uncertain linguistic variables and apply them to solve the UMSDM problems.

The reminder of this paper is shown as follows. In section 1, we give an brief introduction of the research background. In Section 2, we introduced some basic concepts of 2-dimension uncertain linguistic information.In Section 3, we brief review some power aggregation operators based on 2-dimension uncertain linguistic variables. In Section 4, we introduce the basic principles of dynamic interactive group decision making method based on 2-dimension uncertain linguistic variables .In Section 5, we give the decision making steps of dynamic interactive group decision. making method. In Section 6, we give one example to illustrate the decision making steps based on the proposed method .In the last Section , we give the concluding remarks and future research directions.

## 2. Preliminaries： $[s_{a1},s_{b1}] \oplus [s_{a2},s_{b2}] = [s_{a1+a2}, s_{b1+b2}]$

### 2.1 uncertain linguistic variable

When decision makers give the evaluation of the objects, generally it is essential to set an appropriate linguistic assessment in advance .Suppose $\bar{s} = [s_a, s_b]$, $s_a, s_b \in \bar{S}$ 且 $a \leq b$, $s_a, s_b$ are the lower limit and upper limit of $\bar{s}$ respectively, then we call $\bar{s}$ an uncertain linguistic variable. Suppose we use $\bar{S}$ to express the set of all uncertain linguistic variables.

**Definition 1,** Let $\bar{s}_1 = [s_{a1}, s_{b1}]$, $\bar{s}_2 = [s_{a2}, s_{b2}]$ be any two uncertain linguistic variables, then the operational rules are shown as follows:

（1） $\bar{s}_1 \oplus \bar{s}_2 =$ (1)

（2） $\bar{s}_1 \otimes \bar{s}_2 = [s_{a1}, s_{b1}] \otimes [s_{a2}, s_{b2}] = [s_{a1 \times a2}, s_{b1 \times b2}]$ (2)

（3） $\bar{s}_1 / \bar{s}_2 = [s_{a1}, s_{b1}] / [s_{a2}, s_{b2}] = [s_{a1/b2}, s_{b1/a2}]$ (3)

（4） $\lambda \bar{s}_1 = \lambda [s_{a1}, s_{b1}] = [s_{\lambda * a1}, s_{\lambda * b1}]$ (4)

（5） $\lambda(\bar{s}_1 \oplus \bar{s}_2) = \lambda \bar{s}_1 \oplus \lambda \bar{s}_2$ (5)

（6） $(\lambda_1 + \lambda_2)\bar{s}_1 = \lambda_1 \bar{s}_1 \oplus \lambda_2 \bar{s}_2$ (6)

### 2.2 The definition of two-dimension uncertain linguistic variable

**Definition** 2: let $\hat{s} = ([\dot{s}_a, \dot{s}_b], [\ddot{s}_c, \ddot{s}_d])$, where $[\dot{s}_a, \dot{s}_b]$ is I class uncertain linguistic information, which represents decision maker's judgement to evaluated object, and $[\dot{s}_a, \dot{s}_b]$ are the elements from the predefined linguistic assessment set $S_I = (\dot{s}_0, \dot{s}_1, \cdots, \dot{s}_{l-1})$ while $[\ddot{s}_c, \ddot{s}_d]$ is II class uncertain linguistic information, which represents the subjective evaluation on the reliability of their given results. And $[\ddot{s}_c, \ddot{s}_d]$ are the elements from the predefined linguistic assessments set $S_{II} = (\ddot{s}_0, \ddot{s}_1, \cdots, \ddot{s}_{z-1})$, then $\hat{s}$ is called the 2-dimension uncertain linguistic variable。

## 2.3. The operational rules and the characteristic of two dimension uncertain linguistic variables

For any two 2-dimension uncertain linguistic variables $\hat{s}_1 = ([\dot{s}_{a1}, \dot{s}_{b1}], [\ddot{s}_{c1}, \ddot{s}_{d1}])$ and $\hat{s}_2 = ([\dot{s}_{a2}, \dot{s}_{b2}], [\ddot{s}_{c2}, \ddot{s}_{d2}])$, the operational rules are given shown as follows:

(1) $\hat{s}_1 \oplus \hat{s}_2 = ([\dot{s}_{a1}, \dot{s}_{b1}], [\ddot{s}_{c1}, \ddot{s}_{d1}]) \oplus ([\dot{s}_{a2}, \dot{s}_{b2}], [\ddot{s}_{c2}, \ddot{s}_{d2}])$

$$= ([\dot{s}_{a1+a2}, \dot{s}_{b1+b2}], [\ddot{s}_{\min(c1,c2)}, \ddot{s}_{\min(d1,d2)}]) \quad (7)$$

(2) $\hat{s}_1 \otimes \hat{s}_2 = ([\dot{s}_{a1}, \dot{s}_{b1}], [\ddot{s}_{c1}, \ddot{s}_{d1}]) \otimes ([\dot{s}_{a2}, \dot{s}_{b2}], [\ddot{s}_{c2}, \ddot{s}_{d2}])$

$$= ([\dot{s}_{a1 \times a2}, \dot{s}_{b1 \times b2}], [\ddot{s}_{\min(c1,c2)}, \ddot{s}_{\min(d1,d2)}]) \quad (8)$$

(3) $\hat{s}_1 / \hat{s}_2 = ([\dot{s}_{a1}, \dot{s}_{b1}], [\ddot{s}_{c1}, \ddot{s}_{d1}]) / ([\dot{s}_{a2}, \dot{s}_{b2}], [\ddot{s}_{c2}, \ddot{s}_{d2}])$

$$= ([\dot{s}_{a1/b2}, \dot{s}_{b1/a2}], [\ddot{s}_{\min(c1,c2)}, \ddot{s}_{\min(d1,d2)}]), a_2, b_2 \neq 0 \quad (9)$$

(4) $\lambda \hat{s}_1 = \lambda([\dot{s}_{a1}, \dot{s}_{b1}], [\ddot{s}_{c1}, \ddot{s}_{d1}]) = \lambda([\dot{s}_{\lambda \times a1}, \dot{s}_{\lambda \times b1}], [\ddot{s}_{c1}, \ddot{s}_{d1}])(\lambda \geq 0)$ (10)

(5) $(\hat{s}_1^\lambda) = ([\dot{s}_{a1}, \dot{s}_{b1}], [\ddot{s}_{c1}, \ddot{s}_{d1}])^\lambda = \lambda([\dot{s}_{(a1)^\lambda}, \dot{s}_{(b1)^\lambda}], [\ddot{s}_{c1}, \ddot{s}_{d1}])(\lambda \geq 0)$ (11)

Let $\hat{s}_1 = ([\dot{s}_{a1}, \dot{s}_{b1}], [\ddot{s}_{c1}, \ddot{s}_{d1}])$, $\hat{s}_2 = ([\dot{s}_{a2}, \dot{s}_{b2}], [\ddot{s}_{c2}, \ddot{s}_{d2}])$ and $\hat{s}_3 = ([\dot{s}_{a3}, \dot{s}_{b3}], [\ddot{s}_{c3}, \ddot{s}_{d3}])$ be any three 2-dimension uncertain linguistic variables and $\lambda_1, \lambda_2, \lambda_3 \geq 0$.

The two-dimension uncertain linguistic variables satisfy the following properties.

(1) $\hat{s}_1 \oplus \hat{s}_2 = \hat{s}_2 \oplus \hat{s}_1$ (12)

(2) $\hat{s}_1 \otimes \hat{s}_2 = \hat{s}_2 \otimes \hat{s}_1$ (13)

(3) $\hat{s}_1 \oplus \hat{s}_2 \oplus \hat{s}_3 = \hat{s}_1 \oplus (\hat{s}_2 \oplus \hat{s}_3)$ (14)

(4) $\hat{s}_1 \otimes \hat{s}_2 \otimes \hat{s}_3 = \hat{s}_1 \otimes (\hat{s}_2 \otimes \hat{s}_3)$ (15)

(5) $\hat{s}_1 \otimes (\hat{s}_2 \oplus \hat{s}_3) = (\hat{s}_1 \otimes \hat{s}_2) \oplus (\hat{s}_1 \otimes \hat{s}_3)$  (16)

(6) $\lambda(\hat{s}_1 \oplus \hat{s}_2) = (\lambda\hat{s}_1) \oplus (\lambda\hat{s}_2)$  (17)

(7) $(\lambda_1 + \lambda_2)\hat{s}_1 = (\lambda\hat{s}_1) \oplus (\lambda\hat{s}_2)$  (18)

### 2.4. The expectation of the two-dimension uncertain linguistic variable

**Definition 3** Let $\hat{s}_1 = ([\dot{s}_{a1}, \dot{s}_{b1}], [\ddot{s}_{c1}, \ddot{s}_{d1}])$ be a two-dimension uncertain linguistic variable, $[\dot{s}_a, \dot{s}_b]$ are the elements from the predefined linguistic assessment set $S_I = (\dot{s}_0, \dot{s}_1, \cdots, \dot{s}_{l-1})$ and $[\ddot{s}_c, \ddot{s}_d]$ are the elements from the predefined linguistic assessment set $S_{II} = (\ddot{s}_0, \ddot{s}_1, \cdots, \ddot{s}_{t-1})$, then the expectation $E(\hat{s}_1)$ of $\hat{s}_1$ is defined as

$$E(\hat{s}_1) = \frac{a_1 + b_1}{2*(l-1)} * \frac{c_1 + d_1}{2*(z-1)}$$  (19)

### 2.5. The comparison of two-dimension uncertain linguistic variables

Let $\hat{s}_1 = ([\dot{s}_{a1}, \dot{s}_{b1}], [\ddot{s}_{c1}, \ddot{s}_{d1}])$ and $\hat{s}_2 = ([\dot{s}_{a2}, \dot{s}_{b2}], [\ddot{s}_{c2}, \ddot{s}_{d2}])$ be any two-dimension uncertain linguistic variables, if $E(\hat{s}_1) \geq E(\hat{s}_2)$, then $\hat{s}_1 \geq \hat{s}_2$, or vice versa

### 2.6. The distance between two-dimension uncertain linguistic variables

Let $\hat{s}_1 = ([\dot{s}_{a1}, \dot{s}_{b1}], [\ddot{s}_{c1}, \ddot{s}_{d1}])$, $\hat{s}_2 = ([\dot{s}_{a2}, \dot{s}_{b2}], [\ddot{s}_{c2}, \ddot{s}_{d2}])$ be any two-dimension uncertain linguistic variables, and $f: \hat{S} \times \hat{S} \to R$. If $d(\hat{s}_1, \hat{s}_2)$ satisfies the following condition

(1) $0 \leq d(\hat{s}_1, \hat{s}_2) \leq 1, d(\hat{s}_1, \hat{s}_2) = 0$  (20)

(2) $d(\hat{s}_1, \hat{s}_2) = d(\hat{s}_2, \hat{s}_1)$  (21)

(3) $d(\hat{s}_1, \hat{s}_2) + d(\hat{s}_2, \hat{s}_3) \geq d(\hat{s}_1, \hat{s}_3)$  (22)

then $d(\hat{s}_1, \hat{s}_2)$ is the distance between two-dimension uncertain linguistic variables, the Hamming distance of $\hat{s}_1, \hat{s}_2$ is defined as follows:

$$d(\hat{s}_1, \hat{s}_2) = \frac{1}{4(l-1)} \left( \left| a_1 \times \frac{c_1}{z-1} - a_2 \times \frac{c_2}{z-1} \right| + \left| a_1 \times \frac{d_1}{z-1} - a_2 \times \frac{d_2}{z-1} \right| \right.$$

$$\left. + \left| b_1 \times \frac{c_1}{z-1} - b_2 \times \frac{c_2}{z-1} \right| + \left| b_1 \times \frac{d_1}{z-1} - b_2 \times \frac{d_2}{z-1} \right| \right)$$  (23)

## 3. The two-dimension uncertain linguistic generalized aggregation operators

Definition Let $\hat{s}_k = ([\dot{s}_{ak}, \dot{s}_{bk}], [\ddot{s}_{ck}, \ddot{s}_{dk}])$ be the collection of two-dimension uncertain linguistic variables, and $2DULGWA: \Omega^p \to \Omega$. If

$$2DULGWA(\hat{s}_1, \hat{s}_2, \ldots, \hat{s}_p) = \left(\sum_{k=1}^{p} w_k \hat{s}_k^{\alpha}\right)^{1/\alpha} \qquad (24)$$

where $\Omega$ is the set of all two-dimension uncertain li9nguistic variables, and $w = (w_1, w_2, \ldots, w_p)^T$ is the weight vector of $\hat{s}_k (k=1,2,\ldots p)$, satisfying $w_k \geq 0$, $\sum_{k=1}^{p} w_k = 1$. $\alpha$ is a parameter such that $\alpha \in (-\infty, 0) \cup (0, +\infty)$, then 2DULGWA is called the 2-dimensional uncertain linguistic generalized aggregation(2DULGWA) operator based on the operational rules of the 2-dimension uncertain linguistic variables, we have the following theorems.

**Theorem 1** let $\hat{s}_k = ([\dot{s}_{ak}, \dot{s}_{bk}], [\ddot{s}_{ck}, \ddot{s}_{dk}])$ be the collection of two-dimension uncertain linguistic variables, then

$$2DULGWA(\hat{s}_1, \hat{s}_2, \ldots, \hat{s}_p) = \left(\sum_{k=1}^{p} w_k \hat{s}_k^{\alpha}\right)^{1/\alpha} = \left(\left[\dot{s}_{\left(\sum_{k=1}^{p} w_k a_k^{\alpha}\right)^{1/\alpha}}, \dot{s}_{\left(\sum_{k=1}^{p} w_k b_k^{\alpha}\right)^{1/\alpha}}\right], \left[\ddot{s}_{\min_k c_k}, \ddot{s}_{\min_k d_k}\right]\right) \qquad (25)$$

**Theorem2 idempotency :**

Let $\hat{s}_k = \hat{s}, (\hat{s}_k = ([\dot{s}_a, \dot{s}_b], [\ddot{s}_c, \ddot{s}_d]))$, then

$$2DULGWA(\hat{s}_1, \hat{s}_2, \ldots, \hat{s}_p) = \hat{s}$$

**Theorem3 Commutativity :** let $(\hat{s}_1', \hat{s}_2', \ldots, \hat{s}_p')$ be any permutation of $(\hat{s}_1, \hat{s}_2, \ldots, \hat{s}_p)$, then

$$2DULGWA(\hat{s}_1, \hat{s}_2, \ldots, \hat{s}_p) = 2DULGWA(\hat{s}_1', \hat{s}_2', \ldots, \hat{s}_p')$$

**Thoerem4 Monotomicity:** if $\hat{s}_k \leq \hat{s}_k'$ for all k, then

$$2DULGWA(\hat{s}_1, \hat{s}_2, \ldots, \hat{s}_p) \leq 2DULGWA(\hat{s}_1', \hat{s}_2', \ldots, \hat{s}_p')$$

**Theorem5 (Boundedness)**
The 2DULGWA operator lies between the max and min operator, i.e.,

$$\min(\hat{s}_1, \hat{s}_2, \ldots, \hat{s}_p) \leq 2DULGWA(\hat{s}_1, \hat{s}_2, \ldots, \hat{s}_p) \leq \max(\hat{s}_1, \hat{s}_2, \ldots, \hat{s}_p)$$

$$2DULGWA(\hat{s}_1, \hat{s}_2, \ldots, \hat{s}_p) = \sum_{k=1}^{p} w_k \hat{s}_k$$

When $\alpha = 1$, 2DUKGWA operator can be reduced to the 2-dimension uncertain linguistic weighted aggregation（2DULWA）operator.

## 3. An approach to Multi-round Dynamic Group Decision Making Method based on two-dimensional language evaluation information

### 3.1 The description of Multi-round Dynamic Group Decision Making Method based on two-dimensional language evaluation information

Suppose a multiple attribute dynamic interactive group decision making problem with 2-dimension uncertain linguistic information..Let $E = \{e_1, e_2, \cdots e_p\}$ be the set of decision maker in the group decision-making problem and $A = \{A_1, A_2, \cdots, A_m\}$ be the set of alternatives. $\lambda = (\lambda_1, \lambda_2 \cdots \lambda_p)$ is the decision-maker weight vector, satisfying $0 \leq \lambda_k \leq 1, \sum_{k=1}^{p} \lambda_k = 1$. The decision-making process adopts multiple rounds of group decision-making. After t rounds of interaction, the decision-maker's preferences are fully expressed and corrected.In each round of group interaction, the scores vector of the decision-makers obtained are recorded as $s_i^t = (s_{i1}^t, s_{i2}^t, \cdots, s_{in}^t)$, in which $s_{ij}^t$ represents the decision maker $e_k^t$ in the $t$ round for alternative $A_i$ and $s_{ij}^{kt} = ([x_{ij}^{hk}, x_{ij}^{uk}], [g_{ij}^{hk}, g_{ij}^{uk}])$, $(i = 1, 2, \cdots, p; j = 1, 2, \cdots, m; t = 1, 2, \cdots, T)$. As for decision maker $e_i$, after t round interaction, the evaluation scores of the decision makers set constitute a vector set $S = (s_i^1, s_i^2, \cdots, s_i^T)$.

### 3.2 Vector fitting methods

Due to the interaction in the process of group decision in each round ,there are individual preference information adjustment, expert evaluation information each round of the decision-making process of expert evaluation information contains part of real preference information. In order to more accurately obtain expert real preference information, we need calculate the comprehensive value of interaction for each round of the individual preference information contained, So this paper intends to use vector fitting method to obtain the true stable preference vector of experts.

Since the Angle between vectors reflects the degree of approximation between vectors,

the cosine value of the Angle between two vectors is generally calculated to judge the degree of approximation between two vectors.

Cosine of the Angle between two vectors is defined as the following:

$$\cos\alpha_{t1,t2} = \langle s_i^{t1}, s_i^{t2}\rangle / \|s_i^{t1}\| \cdot \|s_i^{t2}\|, t1,t2 \in [1,T], t1 \neq t2$$

where $-1 \leq \cos\alpha \leq 1$, if the greater the value of $\cos\alpha$, the more similar of two vectors $s_i^{t1}, s_i^{t2}$.

Based on the above ideas, to obtain the true and stable decision evaluation vector of the decision-makers $e_i$ for solution set, it can be transformed into the score vector closest to each round of the scoring vector.

Based on the above ideas, in order to obtain the real and stable decision evaluation vector of the decision-maker $e_i$ for the scheme set $A = \{A_1, A_2, \cdots, A_m\}$, it can be converted into calculate the evaluation vector $\hat{s}_i = (\hat{s}_{i1}, \hat{s}_{i2}, \cdots, \hat{s}_{in})$ which is closest to the evaluation vector of each round $s_i^t = (s_{i1}^t, s_{i2}^t, \cdots, s_{in}^t)$.

If we normalize $s_i^t$, then $\|s_i^{t1}\|=1, \|s_i^{t2}\|=1, \|s_i^t\|=1$. and the value of $\cos\alpha_{t1,t2}$ will be only influenced by $\langle s_i^{t1}, s_i^{t2}\rangle$.

In order to calculate the closest evaluation vector $\hat{s}_i = (\hat{s}_{i1}, \hat{s}_{i2}, \cdots, \hat{s}_{in})$, it can be solved by the following optimization problem:

$$\max \sum_{t=1}^{T} \langle s_i^t \cdot s_i{'}\rangle = \max \sum_{t=1}^{T} s_i^t \cdot s_i{'} = \max \sum_{t=1}^{T} (s_i^t \cdot s_i{'})^2$$

s.t. $\|s_i\| = 1$ 即 $s_i s_i{'} = 1$.

Solution:

Let $f = \sum_{t=1}^{T}(s_i^t \cdot s_i{'})^2 = \sum_{t=1}^{T}(s_i^t \cdot s_i{'} \cdot s_i^t \cdot s_i{'}) = s_i{'}\sum_{t=1}^{T}[(s_i^t)\cdot s_i^t]s_i{'}$

Suppose $F = \sum_{i=1}^{t}(s_i^t)s_i^t = S_i \cdot S_i^T$,

where $S_i = \begin{bmatrix} s_{i1}^1 & s_{i1}^1 & \cdots & s_{in}^1 \\ s_{i1}^2 & s_{i2}^2 & \cdots & s_{in}^2 \\ \vdots & \vdots & \ddots & \vdots \\ s_{i1}^T & s_{i2}^T & \cdots & s_{in}^T \end{bmatrix}$

Then $f = s_i F s_i{}'$

Lagrange multiplier function is constructed by Lagrange multiplier method as following:

$$g(s_{i1}, s_{i2}, \cdots, s_{in}, \lambda) = s_i F s_i{}' - \lambda(s_i s_i{}' - 1)$$

Let $\dfrac{\partial g(s_{i1}, s_{i2}, \ldots s_{in}, \lambda)}{\partial s_{ij}} = 0 \quad j = 1, 2, \ldots, n$

$\dfrac{\partial g(s_{i1}, s_{i2}, \ldots s_{in}, \lambda)}{\partial \lambda} = 0$

Then $F s_i{}' = \lambda s_i{}'$

Because $s_i s_i{}' = \|s_i\| = 1$, then $\lambda$ is the eigenvalue if the matrix $F$, $s_i{}'$ is its corresponding eigenvector. $s_i{}'$ is the evaluation vector closest to each evaluation vector. In this way, Calculating the real and stable preference information of the expert is transformed into the maximum eigenvector corresponding to the maximum eigenvalue of the matrix. The vector $\hat{s}_i = (\hat{s}_{i1}, \hat{s}_{i2}, \cdots, \hat{s}_{in})$ not only consider individual preference information, but also the interactive process and preference adjustment process of group decision.

### 3.3 Expert weight based on uncertainty

**Definition 1:** Let $\hat{s}_1 = ([\dot{s}_{a1}, \dot{s}_{b1}], [\ddot{s}_{c1}, \ddot{s}_{d1}])$ be a two-dimension uncertain linguistic variable, the uncertainty degree $\beta(s_k)$ is defined as:

$$\beta(s_k) = \dfrac{1}{2}\left(\dfrac{b1-a1}{l}\right) + \left(\dfrac{d1-c1}{t}\right) \qquad (4\text{-}1)$$

**Definition 2: Uncertainty degree of individual decision matrix given by experts is defined as the following:**

$$\beta(s_i) = \sum_{t=1}^{T} \sum_{j=1}^{m} \beta(s_{ij}^t), i = 1, \cdots T, j = 1, \cdots m$$

**Definition 3**

The lower the uncertainty of the individual decision matrix given by the expert, the more accurate the expert's decision is, and the greater the weight should be assigned to the expert, and vice the smaller the weight should be assigned to the expert. Thus, the calculation formula of the expert based on the uncertainty is given as follows:

$$\lambda_i^{(1)} = \frac{\frac{1}{\beta(s_i)}}{\sum_{i=1}^{p} \frac{1}{\beta(s_i)}} \quad (4\text{-}3)$$

### 3.4 Objective weight of experts based on deviation degree

Due to group decision making is a process of multiple experts consultation, the final results reflect the opinions of the experts of compromise, should be consistent, so you can according to the experts to the deviation between the individual decision and group decision expert weight is determined, and the greater the difference of group decision expert, together with the group intend to deviate from the farther, smaller weights should be assigned to abate the influence of the expert group decision results, whereas the smaller deviation, indicates that the expert decision-making and group decision-making of the more consistent, should be assigned to its greater weight, in order to improve the degree of expert group's consensus, specific calculation method is as follows:：

Definition 1:

The deviation degree $\zeta_k$ between individual experts matrix and group decision matrix ,which is calculated by using the distance formula:

$$\zeta_i = \sum_{t=1}^{T} \sum_{j=1}^{m} d(s_{ij}^t, y_{ij}) \quad (4\text{-}4)$$

Definition 2:

Objective weight of experts based on deviation degree $\lambda_i^{(2)}$ is given as follows:

$$\lambda_i^{(2)} = \frac{1}{\zeta_i} \Big/ \sum_{i=1}^{p} \frac{1}{\zeta_i} \quad (4\text{-}5)$$

### 3.5 Expert comprehensive objective weight

The expert weight based on uncertainty reflects the importance of experts by comparing the accuracy of each expert's decision information, while the expert weight based on deviation reflects the importance of experts by calculating the consistency between the expert's decision information and group decision information. In order to obtain more comprehensive information of expert weight, decision makers can set parameters $\eta(0 \leq \eta \leq 1)$ according to specific decision problems By aggregating two kinds of weights, the comprehensive objective weight of experts can be obtained as

follows:

$$\lambda_i = \eta \lambda_i^{(1)} + (1-\eta)\lambda_i^{(2)} \qquad (4\text{-}6)$$

## 4. Decision making step based on two-dimensional language evaluation information

**Step 1: calculate the uncertainty degree** $\beta(s_k)$

$$\beta(s_k) = \frac{1}{2}\left(\frac{b1-a1}{l}\right) + \left(\frac{d1-c1}{t}\right)$$

**Step 2: calculate the Expert weight based on uncertainty degree** $\lambda_i^{(1)}$

$$\beta(s_i) = \sum_{t=1}^{T}\sum_{j=1}^{m} \beta(s_{ij}^t) \quad i=1,\cdots\cdots T, j=1,\cdots\cdots m$$

$$\lambda_i^{(1)} = \frac{\dfrac{1}{\beta(s_i)}}{\sum_{i=1}^{p}\dfrac{1}{\beta(s_i)}}$$

**Step 3: calculate the comprehensive evaluation value of each alternative** $\hat{Y} = [\hat{y}_{ij}]_{m\times n}$,

$$\hat{y}_{ij}^t = 2DULGWA(\hat{s}_1, \hat{s}_2, \ldots, \hat{s}_p) = \left(\sum_{k=1}^{p} w_k \hat{s}_k^{\alpha}\right)^{1/\alpha} = \left(\left[\dot{s}_{\left(\sum_{k=1}^{p} w_k a_k^{\alpha}\right)^{1/\alpha}}, \dot{s}_{\left(\sum_{k=1}^{p} w_k b_k^{\alpha}\right)^{1/\alpha}}\right], \left[\ddot{s}_{\min_k c_k}, \ddot{s}_{\min_k d_k}\right]\right)$$

The group decision matrix $\hat{Y} = [\hat{y}_{ij}]_{m\times n}$ is calculated by the 2DULGWA operator, where

$$\hat{Y} = \begin{bmatrix} \hat{y}_{11} & \hat{y}_{12} & \cdots & \hat{y}_{1n} \\ \hat{y}_{21} & \hat{y}_{22} & \cdots & \hat{y}_{2n} \\ \vdots & \vdots & \ddots & \vdots \\ \hat{y}_{m1} & \hat{y}_{m2} & \cdots & \hat{y}_{mn} \end{bmatrix} \quad \text{and} \quad \hat{y}_{ij} = \left([x_{ij}^H, x_{ij}^U],[g_{ij}^H, g_{ij}^U]\right)$$

**Step 4: calculate the distance matrix between the** individual experts matrix and group decision matrix $d(\hat{s}_1, \hat{s}_2)$

$$d(\hat{s}_1, \hat{s}_2) = \frac{1}{4(l-1)}\left(\left|a_1 \times \frac{c_1}{z-1} - a_2 \times \frac{c_2}{z-1}\right| + \left|a_1 \times \frac{d_1}{z-1} - a_2 \times \frac{d_2}{z-1}\right|\right.$$

$$+\left|b_1 \times \frac{c_1}{z-1} - b_2 \times \frac{c_2}{z-1}\right| + \left|b_1 \times \frac{d_1}{z-1} - b_2 \times \frac{d_2}{z-1}\right|\right)$$

Step 5: calculate the experts weights based on deviation degree $\lambda_i^{(2)}$

$$\zeta_i = \sum_{t=1}^{T}\sum_{j=1}^{m} d(s_{ij}^t, y_{ij})$$

$$\lambda_i^{(2)} = \frac{1}{\zeta_i} \Big/ \sum_{i=1}^{p} \frac{1}{\zeta_i}$$

**Step6: calculate Experts' comprehensive objective weight** $\lambda_i$

$$\lambda_i = \eta \lambda_i^{(1)} + (1-\eta)\lambda_i^{(2)}, \text{where } \eta(0 \le \eta \le 1)$$

**step7: calculate the expectation** $E(\hat{s}_{ij}^t)$ associated with two-dimension uncertain linguistic variable

$$E(\hat{s}_{ij}^t) = \frac{a_{ij}^t + b_{ij}^t}{2*(l-1)} * \frac{c_{ij}^t + d_{ij}^t}{2*(z-1)}$$

**Step 8: According to the expectation value** $E(\hat{s}_{ij}^t)$, **obtain the comprehensive** evaluation vector of all round decision with one decision-maker $s_i^t = (E(s_{i1}^t), E(s_{i2}^t), \ldots, E(s_{im}^t))$ and the all decision-makers matrix is as following:

$$S_i = \begin{bmatrix} E(s_{i1}^1) & E(s_{i2}^1) & \cdots & E(s_{im}^1) \\ E(s_{i1}^2) & E(s_{i2}^2) & \cdots & E(s_{im}^2) \\ \vdots & \vdots & \ddots & \vdots \\ E(s_{i1}^T) & E(s_{i2}^T) & \cdots & E(s_{in}^T) \end{bmatrix}$$

**Step 9**: calculate the evaluation vector $\hat{s}_i = (\hat{s}_{i1}, \hat{s}_{i2}, \cdots, \hat{s}_{in})$ which is closest to the evaluation vector of each round $s_i^t = (s_{i1}^t, s_{i2}^t, \cdots, s_{in}^t)$.

**Step10**: rank all the alternatives and select the best ones according to $\hat{s}_g$

$$\hat{s}_g = \left(\sum_{i=1}^{p} \lambda_i \hat{s}_{i1}, \sum_{i=1}^{p} \lambda_i \hat{s}_{i2}, \ldots, \sum_{i=1}^{p} \lambda_i \hat{s}_{im}\right)$$

# 4. Illustrative example

In this section, we will provide one example. One is for evaluation for the supply chain and byt this example, we can explain the actual application of the proposed method.

This is an example of supply chain project innovation ability evaluation. There are five supply chains $\{a_1, a_2, a_3, a_4, a_5\}$, Four experts $\{e_1, e_2, e_3, e_4\}$ were invited to evaluate these supply chain project. The evaluation values given by the experts take the form of two-dimension uncertain linguistic variables. The experts adopted I class linguistic set $S_I = (\dot{s}_0, \dot{s}_1, \dot{s}_2, \dot{s}_3, \dot{s}_4, \dot{s}_5, \dot{s}_6)$ and the II class linguistic set $S_{II} = (\ddot{s}_0, \ddot{s}_1, \ddot{s}_2, \ddot{s}_3, \ddot{s}_4)$. After three rounds of interaction, the rating vectors of the schemes given by decision makers are shown in table 6-1, 6-2, 6-3

Table 1 the supply chain project evaluation value given by experts in the first round

| Expert | Project $(a_1)$ | Project $(a_2)$ | Project $(a_3)$ | Project $(a_4)$ | Project $(a_5)$ |
|---|---|---|---|---|---|
| $e_1$ | $([\dot{s}_5,\dot{s}_5],[\ddot{s}_2,\ddot{s}_3])$ | $([\dot{s}_2,\dot{s}_3],[\ddot{s}_3,\ddot{s}_3])$ | $([\dot{s}_4,\dot{s}_5],[\ddot{s}_4,\ddot{s}_4])$ | $([\dot{s}_3,\dot{s}_4],[\ddot{s}_1,\ddot{s}_2])$ | $([\dot{s}_5,\dot{s}_3],[\ddot{s}_4,\ddot{s}_3])$ |
| $e_2$ | $([\dot{s}_3,\dot{s}_4],[\ddot{s}_2,\ddot{s}_3])$ | $([\dot{s}_5,\dot{s}_5],[\ddot{s}_3,\ddot{s}_3])$ | $([\dot{s}_3,\dot{s}_3],[\ddot{s}_4,\ddot{s}_4])$ | $([\dot{s}_4,\dot{s}_4],[\ddot{s}_1,\ddot{s}_2])$ | $([\dot{s}_4,\dot{s}_2],[\ddot{s}_4,\ddot{s}_3])$ |
| $e_3$ | $([\dot{s}_2,\dot{s}_3],[\ddot{s}_2,\ddot{s}_3])$ | $([\dot{s}_3,\dot{s}_4],[\ddot{s}_3,\ddot{s}_3])$ | $([\dot{s}_3,\dot{s}_4],[\ddot{s}_4,\ddot{s}_4])$ | $([\dot{s}_4,\dot{s}_5],[\ddot{s}_1,\ddot{s}_2])$ | $([\dot{s}_3,\dot{s}_4],[\ddot{s}_4,\ddot{s}_3])$ |
| $e_4$ | $([\dot{s}_5,\dot{s}_6],[\ddot{s}_2,\ddot{s}_3])$ | $([\dot{s}_1,\dot{s}_2],[\ddot{s}_3,\ddot{s}_3])$ | $([\dot{s}_2,\dot{s}_3],[\ddot{s}_4,\ddot{s}_4])$ | $([\dot{s}_3,\dot{s}_4],[\ddot{s}_1,\ddot{s}_2])$ | $([\dot{s}_5,\dot{s}_3],[\ddot{s}_4,\ddot{s}_3])$ |

Table 2 the supply chain project evaluation value given by experts in the second round

| Expert | Project $(a_1)$ | Project $(a_2)$ | Project $(a_3)$ | Project $(a_4)$ | Project $(a_5)$ |
|---|---|---|---|---|---|
| $e_1$ | $([\dot{s}_4,\dot{s}_4],[\ddot{s}_3,\ddot{s}_4])$ | $([\dot{s}_3,\dot{s}_4],[\ddot{s}_2,\ddot{s}_3])$ | $([\dot{s}_3,\dot{s}_4],[\ddot{s}_3,\ddot{s}_3])$ | $([\dot{s}_5,\dot{s}_6],[\ddot{s}_3,\ddot{s}_4])$ | $([\dot{s}_6,\dot{s}_4],[\ddot{s}_4,\ddot{s}_4])$ |
| $e_2$ | $([\dot{s}_4,\dot{s}_5],[\ddot{s}_3,\ddot{s}_4])$ | $([\dot{s}_2,\dot{s}_3],[\ddot{s}_2,\ddot{s}_3])$ | $([\dot{s}_4,\dot{s}_5],[\ddot{s}_3,\ddot{s}_3])$ | $([\dot{s}_2,\dot{s}_3],[\ddot{s}_3,\ddot{s}_4])$ | $([\dot{s}_5,\dot{s}_4],[\ddot{s}_4,\ddot{s}_4])$ |
| $e_3$ | $([\dot{s}_3,\dot{s}_4],[\ddot{s}_3,\ddot{s}_4])$ | $([\dot{s}_4,\dot{s}_4],[\ddot{s}_2,\ddot{s}_3])$ | $([\dot{s}_2,\dot{s}_3],[\ddot{s}_3,\ddot{s}_3])$ | $([\dot{s}_3,\dot{s}_4],[\ddot{s}_3,\ddot{s}_4])$ | $([\dot{s}_4,\dot{s}_2],[\ddot{s}_4,\ddot{s}_3])$ |
| $e_4$ | $([\dot{s}_5,\dot{s}_5],[\ddot{s}_3,\ddot{s}_4])$ | $([\dot{s}_4,\dot{s}_5],[\ddot{s}_3,\ddot{s}_3])$ | $([\dot{s}_1,\dot{s}_2],[\ddot{s}_3,\ddot{s}_3])$ | $([\dot{s}_4,\dot{s}_4],[\ddot{s}_3,\ddot{s}_4])$ | $([\dot{s}_5,\dot{s}_5],[\ddot{s}_3,\ddot{s}_4])$ |

Table 3 the supply chain project evaluation value given by experts in the third round

| Expert | Project $(a_1)$ | Project $(a_2)$ | Project $(a_3)$ | Project $(a_4)$ | Project $(a_5)$ |
|---|---|---|---|---|---|
| $e_1$ | $([\dot{s}_4,\dot{s}_4],[\ddot{s}_3,\ddot{s}_4])$ | $([\dot{s}_3,\dot{s}_4],[\ddot{s}_2,\ddot{s}_3])$ | $([\dot{s}_3,\dot{s}_4],[\ddot{s}_3,\ddot{s}_3])$ | $([\dot{s}_5,\dot{s}_6],[\ddot{s}_3,\ddot{s}_4])$ | $([\dot{s}_6,\dot{s}_4],[\ddot{s}_4,\ddot{s}_4])$ |
| $e_2$ | $([\dot{s}_4,\dot{s}_5],[\ddot{s}_3,\ddot{s}_4])$ | $([\dot{s}_2,\dot{s}_3],[\ddot{s}_2,\ddot{s}_3])$ | $([\dot{s}_4,\dot{s}_5],[\ddot{s}_3,\ddot{s}_3])$ | $([\dot{s}_2,\dot{s}_3],[\ddot{s}_3,\ddot{s}_4])$ | $([\dot{s}_5,\dot{s}_4],[\ddot{s}_4,\ddot{s}_4])$ |
| $e_3$ | $([\dot{s}_3,\dot{s}_4],[\ddot{s}_3,\ddot{s}_4])$ | $([\dot{s}_4,\dot{s}_4],[\ddot{s}_2,\ddot{s}_3])$ | $([\dot{s}_2,\dot{s}_3],[\ddot{s}_3,\ddot{s}_3])$ | $([\dot{s}_3,\dot{s}_4],[\ddot{s}_3,\ddot{s}_4])$ | $([\dot{s}_4,\dot{s}_2],[\ddot{s}_4,\ddot{s}_3])$ |
| $e_4$ | $([\dot{s}_5,\dot{s}_5],[\ddot{s}_3,\ddot{s}_4])$ | $([\dot{s}_4,\dot{s}_5],[\ddot{s}_3,\ddot{s}_3])$ | $([\dot{s}_1,\dot{s}_2],[\ddot{s}_3,\ddot{s}_3])$ | $([\dot{s}_4,\dot{s}_4],[\ddot{s}_3,\ddot{s}_4])$ | $([\dot{s}_5,\dot{s}_5],[\ddot{s}_3,\ddot{s}_4])$ |

**Step1: : calculate the** uncertainty degree based on the formula,

$$\beta(s_k) = \frac{1}{2}\left(\frac{b1-a1}{l}\right) + \left(\frac{d1-c1}{t}\right)$$

.

Table 4  The uncertainty degree values in the first round

| Expert | Project $(a_1)$ | Project $(a_2)$ | Project $(a_3)$ | Project $(a_4)$ | Project $(a_5)$ |
|---|---|---|---|---|---|
| $e_1$ | $\beta^1_{11}$=0.125 | $\beta^1_{12}$=0.803 | $\beta^1_{13}$=0.803 | $\beta^1_{14}$=0.208 | $\beta^1_{15}$=0.083 |
| $e_2$ | $\beta^1_{21}$=0.208 | $\beta^1_{22}$=0.000 | $\beta^1_{23}$=0.000 | $\beta^1_{24}$=0.125 | $\beta^1_{25}$=0.000 |
| $e_3$ | $\beta^1_{31}$=0.208 | $\beta^1_{32}$=0.083 | $\beta^1_{33}$=0.083 | $\beta^1_{34}$=0.208 | $\beta^1_{35}$=0.250 |
| $e_4$ | $\beta^1_{41}$=0.208 | $\beta^1_{42}$=0.083 | $\beta^1_{43}$=0.083 | $\beta^1_{44}$=0.208 | $\beta^1_{45}$=0.083 |

Table 5  The uncertainty degree values in the 2-nd round

| Expert | Project $(a_1)$ | Project $(a_2)$ | Project $(a_3)$ | Project $(a_4)$ | Project $(a_5)$ |
|---|---|---|---|---|---|
| $e_1$ | $\beta^1_{11}$=0.125 | $\beta^1_{12}$=0.208 | $\beta^1_{13}$=0.803 | $\beta^1_{14}$=0.208 | $\beta^1_{15}$=0.083 |
| $e_2$ | $\beta^1_{21}$=0.208 | $\beta^1_{22}$=0.208 | $\beta^1_{23}$=0.083 | $\beta^1_{24}$=0.208 | $\beta^1_{25}$=0.083 |
| $e_3$ | $\beta^1_{31}$=0.208 | $\beta^1_{32}$=0.125 | $\beta^1_{33}$=0.083 | $\beta^1_{34}$=0.208 | $\beta^1_{35}$=0.333 |
| $e_4$ | $\beta^1_{41}$=0.125 | $\beta^1_{42}$=0.208 | $\beta^1_{43}$=0.083 | $\beta^1_{44}$=0.125 | $\beta^1_{45}$=0.083 |

Table 6  The uncertainty degree values in the 3-rd round

| Expert | Project $(a_1)$ | Project $(a_2)$ | Project $(a_3)$ | Project $(a_4)$ | Project $(a_5)$ |
|---|---|---|---|---|---|
| $e_1$ | $\beta^1_{11}$=0.125 | $\beta^1_{12}$=0.000 | $\beta^1_{13}$=0.125 | $\beta^1_{14}$=0.083 | $\beta^1_{15}$=0.083 |
| $e_2$ | $\beta^1_{21}$=0.125 | $\beta^1_{22}$=0.083 | $\beta^1_{23}$=0.208 | $\beta^1_{24}$=0.000 | $\beta^1_{25}$=0.000 |
| $e_3$ | $\beta^1_{31}$=0.208 | $\beta^1_{32}$=0.000 | $\beta^1_{33}$=0.125 | $\beta^1_{34}$=0.000 | $\beta^1_{35}$=0.333 |
| $e_4$ | $\beta^1_{41}$=0.208 | $\beta^1_{42}$=0.083 | $\beta^1_{43}$=0.208 | $\beta^1_{44}$=0.083 | $\beta^1_{45}$=0.083 |

**Step 2: calculate the Expert weight based on uncertainty degree** $\lambda_i^{(1)}$

$$\lambda_{e_1}^{(1)} = 0.272 \quad \lambda_{e_2}^{(1)} = 0.301$$
$$\lambda_{e_3}^{(1)} = 0.189 \quad \lambda_{e_4}^{(1)} = 0.237$$

Step 3: calculate the comprehensive evaluation value of each alternative $\hat{Y} = [\hat{y}_{ij}]_{m \times n}$,

Table 7  the comprehensive evaluation value of each alternative

| Expert | Project $(a_1)$ | Project $(a_2)$ | Project $(a_3)$ | Project $(a_4)$ | Project $(a_5)$ |
|---|---|---|---|---|---|
| $e_1$ | $([\dot{s}_{4.667}, \dot{s}_{4.667}], [\ddot{s}_2, \ddot{s}_3])$ | $([\dot{s}_{2.667}, \dot{s}_{3.333}], [\ddot{s}_2, \ddot{s}_2])$ | $([\dot{s}_{3.667}, \dot{s}_{4.333}], [\ddot{s}_3, \ddot{s}_3])$ | $([\dot{s}_{4.000}, \dot{s}_{5.000}], [\ddot{s}_1, \ddot{s}_1])$ | $([\dot{s}_{4.000}, \dot{s}_{5.000}], [\ddot{s}_3, \ddot{s}_3])$ |
| $e_2$ | $([\dot{s}_{3.667}, \dot{s}_{4.333}], [\ddot{s}_2, \ddot{s}_3])$ | $([\dot{s}_{3.667}, \dot{s}_{4.333}], [\ddot{s}_2, \ddot{s}_2])$ | $([\dot{s}_{2.667}, \dot{s}_{3.333}], [\ddot{s}_3, \ddot{s}_3])$ | $([\dot{s}_{3.000}, \dot{s}_{3.333}], [\ddot{s}_1, \ddot{s}_1])$ | $([\dot{s}_{5.000}, \dot{s}_{5.333}], [\ddot{s}_3, \ddot{s}_3])$ |
| $e_3$ | $([\dot{s}_{2.667}, \dot{s}_{3.667}], [\ddot{s}_2, \ddot{s}_3])$ | $([\dot{s}_{4.000}, \dot{s}_{4.333}], [\ddot{s}_2, \ddot{s}_2])$ | $([\dot{s}_{2.000}, \dot{s}_{2.667}], [\ddot{s}_3, \ddot{s}_3])$ | $([\dot{s}_{3.667}, \dot{s}_{4.333}], [\ddot{s}_1, \ddot{s}_1])$ | $([\dot{s}_{3.667}, \dot{s}_{4.333}], [\ddot{s}_2, \ddot{s}_3])$ |
| $e_4$ | $([\dot{s}_{4.000}, \dot{s}_{4.667}], [\ddot{s}_2, \ddot{s}_3])$ | $([\dot{s}_{2.333}, \dot{s}_{3.333}], [\ddot{s}_2, \ddot{s}_2])$ | $([\dot{s}_{2.333}, \dot{s}_{3.333}], [\ddot{s}_3, \ddot{s}_3])$ | $([\dot{s}_{3.667}, \dot{s}_{4.333}], [\ddot{s}_1, \ddot{s}_1])$ | $([\dot{s}_{3.000}, \dot{s}_{4.000}], [\ddot{s}_3, \ddot{s}_3])$ |

**Step 4:** calculate the distance matrix between the individual experts matrix and group decision matrix $d(\hat{s}_1, \hat{s}_2)$

Table 8 **the distance matrix between the** individual experts and group decision in the 1-st round

| Expert | Project $(a_1)$ | Project $(a_2)$ | Project $(a_3)$ | Project $(a_4)$ | Project $(a_5)$ |
|---|---|---|---|---|---|
| $e_1$ | $d_{11}^1$=0.035 | $d_{12}^1$=0.063 | $d_{13}^1$=0.250 | $d_{14}^1$=0.073 | $d_{15}^1$=0.000 |
| $e_2$ | $d_{21}^1$=0.052 | $d_{22}^1$=0.292 | $d_{23}^1$=0.125 | $d_{24}^1$=0.118 | $d_{25}^1$=0.188 |
| $e_3$ | $d_{31}^1$=0.069 | $d_{32}^1$=0.090 | $d_{33}^1$=0.292 | $d_{34}^1$=0.115 | $d_{35}^1$=0.167 |
| $e_4$ | $d_{41}^1$=0.122 | $d_{42}^1$=0.049 | $d_{43}^1$=0.063 | $d_{44}^1$=0.073 | $d_{45}^1$=0.000 |

Table 9 **the distance matrix between the** individual experts and group decision in the 1-st round

| Expert | Project $(a_1)$ | Project $(a_2)$ | Project $(a_3)$ | Project $(a_4)$ | Project $(a_5)$ |
|---|---|---|---|---|---|
| $e_1$ | $d_{11}^2$=0.097 | $d_{12}^2$=0.115 | $d_{13}^2$=0.063 | $d_{14}^2$=0.615 | $d_{15}^2$=0.000 |
| $e_2$ | $d_{21}^2$=0.240 | $d_{22}^2$=0.080 | $d_{23}^2$=0.188 | $d_{24}^2$=0.233 | $d_{25}^2$=0.042 |
| $e_3$ | $d_{31}^2$=0.181 | $d_{32}^2$=0.083 | $d_{33}^2$=0.021 | $d_{34}^2$=0.344 | $d_{35}^2$=0.240 |
| $e_4$ | $d_{41}^2$=0.284 | $d_{42}^2$=0.233 | $d_{43}^2$=0.167 | $d_{44}^2$=0.417 | $d_{45}^2$=0.000 |

Table 10 **the distance matrix between the** individual experts and group decision in the 1-st round

| Expert | Project $(a_1)$ | Project $(a_2)$ | Project $(a_3)$ | Project $(a_4)$ | Project $(a_5)$ |
|---|---|---|---|---|---|
| $e_1$ | $d_{11}^3$=0.035 | $d_{12}^3$=0.028 | $d_{13}^3$=0.104 | $d_{14}^3$=0.000 | $d_{15}^3$=0.000 |
| $e_2$ | $d^3_{21}=0.035$ | $d^3_{22}=0.042$ | $d^3_{23}=0.156$ | $d^3_{24}=0.007$ | $d^3_{25}=0.188$ |
| $e_3$ | $d^3_{31}=0.035$ | $d^3_{32}=0.069$ | $d^3_{33}=0.146$ | $d^3_{34}=0.014$ | $d^3_{35}=0.052$ |
| $e_4$ | $d^3_{41}=0.191$ | $d^3_{42}=0.028$ | $d^3_{43}=0.302$ | $d^3_{44}=0.021$ | $d^3_{45}=0.000$ |

Step 5: calculate the experts weights based on deviation degree

$\lambda^{(2)}_{e_1} = 0.306 \quad \lambda^{(2)}_{e_2} = 0.227$
$\lambda^{(2)}_{e_3} = 0.235 \quad \lambda^{(2)}_{e_4} = 0.232$

**step6: calculate Experts' comprehensive objective weight** $\lambda_i$

$\lambda_{e_1} = 0.292 \quad \lambda_{e_2} = 0.257$
$\lambda_{e_3} = 0.217 \quad \lambda_{e_4} = 0.234$

**Step7: calculate the expectation** $E(\hat{s}^t_{ij})$ associated with two-dimension uncertain linguistic variable.

Table 11: the expectation results with respect to experts in the first round

| Expert | Project $(a_1)$ | Project $(a_2)$ | Project $(a_3)$ | Project $(a_4)$ | Project $(a_5)$ |
|---|---|---|---|---|---|
| $e_1$ | $E(s^1_{11})=0.521$ | $E(s^1_{12})=0.313$ | $E(s^1_{13})=0.750$ | $E(s^1_{14})=0.563$ | $E(s^1_{15})=0.563$ |
| $e_2$ | $E(s^1_{21})=0.365$ | $E(s^1_{22})=0.625$ | $E(s^1_{23})=0.500$ | $E(s^1_{24})=0.833$ | $E(s^1_{25})=0.833$ |
| $e_3$ | $E(s^1_{31})=0.260$ | $E(s^1_{32})=0.438$ | $E(s^1_{33})=0.583$ | $E(s^1_{34})=0.583$ | $E(s^1_{35})=0.583$ |
| $e_4$ | $E(s^1_{41})=0.573$ | $E(s^1_{42})=0.188$ | $E(s^1_{43})=0.417$ | $E(s^1_{44})=0.438$ | $E(s^1_{45})=0.438$ |

Table 12 : the expectation results with respect to experts in the first round

| Expert | Project $(a_1)$ | Project $(a_2)$ | Project $(a_3)$ | Project $(a_4)$ | Project $(a_5)$ |
|---|---|---|---|---|---|
| $e_1$ | $E(s^2_{11})=0.583$ | $E(s^2_{12})=0.365$ | $E(s^2_{13})=0.438$ | $E(s^2_{14})=0.802$ | $E(s^2_{15})=0.563$ |
| $e_2$ | $E(s^2_{21})=0.656$ | $E(s^2_{22})=0.260$ | $E(s^2_{23})=0.563$ | $E(s^2_{24})=0.365$ | $E(s^2_{25})=0.688$ |
| $e_3$ | $E(s^2_{31})=0.510$ | $E(s^2_{32})=0.417$ | $E(s^2_{33})=0.313$ | $E(s^2_{34})=0.510$ | $E(s^2_{35})=0.656$ |
| $e_4$ | $E(s^2_{41})=0.729$ | $E(s^2_{42})=0.469$ | $E(s^2_{43})=0.188$ | $E(s^2_{44})=0.583$ | $E(s^2_{45})=0.438$ |

Table 13: the expectation results with respect to experts in the first round

| Expert | Project $(a_1)$ | Project $(a_2)$ | Project $(a_3)$ | Project $(a_4)$ | Project $(a_5)$ |
|---|---|---|---|---|---|
| $e_1$ | $E(s^3_{11})=0.521$ | $E(s^3_{12})=0.250$ | $E(s^3_{13})=0.583$ | $E(s^3_{14})=0.188$ | $E(s^3_{15})=0.563$ |
| $e_2$ | $E(s_{21}^3)$=0.417 | $E(s_{22}^3)$=0.375 | $E(s_{23}^3)$=0.219 | $E(s_{24}^3)$=0.125 | $E(s_{25}^3)$=0.833 |
| $e_3$ | $E(s_{31}^3)$=0.365 | $E(s_{32}^3)$=0.417 | $E(s_{33}^3)$=0.146 | $E(s_{34}^3)$=0.167 | $E(s_{35}^3)$=0.365 |
| $e_4$ | $E(s_{41}^3)$=0.260 | $E(s_{42}^4)$=0.208 | $E(s_{43}^3)$=0.656 | $E(s_{44}^3)$=0.188 | $E(s_{45}^3)$=0.438 |

**Step 8: According to the expectation value $E(\hat{s}_{ij}^t)$, obtain the comprehensive evaluation vector of all decision maker**

$s_1^1 = (0.521, 0.313, 0.750, 0.563, 0.563), s_1^2 = (0.583, 0.365, 0.438, 0.802, 0.563), s_1^3 = (0.521, 0.250, 0.583, 0.188, 0.563)$

$s_2^1 = (0.365, 0.625, 0.500, 0.833, 0.833), s_2^2 = (0.656, 0.260, 0.563, 0.365, 0.688), s_2^3 = (0.417, 0.375, 0.219, 0.125, 0.833)$

$s_3^1 = (0.260, 0.438, 0.583, 0.583, 0.583), s_3^2 = (0.510, 0.417, 0.313, 0.510, 0.656), s_3^3 = (0.365, 0.417, 0.146, 0.167, 0.365)$

$s_4^1 = (0.573, 0.188, 0.417, 0.438, 0.438), s_4^2 = (0.729, 0.469, 0.188, 0.583, 0.438), s_4^3 = (0.260, 0.208, 0.656, 0.188, 0.438)$

the all decision-makers matrix is as following:

$$S_1 = \begin{pmatrix} 0.520 & 0.313 & 0.750 & 0.563 & 0.563 \\ 0.583 & 0.365 & 0.438 & 0.802 & 0.563 \\ 0.521 & 0.250 & 0.583 & 0.188 & 0.563 \end{pmatrix}$$

$$S_2 = \begin{pmatrix} 0.365 & 0.625 & 0.500 & 0.833 & 0.833 \\ 0.656 & 0.260 & 0.563 & 0.365 & 0.688 \\ 0.417 & 0.375 & 0.219 & 0.125 & 0.833 \end{pmatrix}$$

$$S_3 = \begin{pmatrix} 0.260 & 0.438 & 0.583 & 0.583 & 0.583 \\ 0.510 & 0.417 & 0.313 & 0.510 & 0.656 \\ 0.365 & 0.417 & 0.146 & 0.167 & 0.365 \end{pmatrix}$$

$$S_4 = \begin{pmatrix} 0.573 & 0.188 & 0.417 & 0.438 & 0.438 \\ 0.729 & 0.469 & 0.188 & 0.583 & 0.438 \\ 0.260 & 0.208 & 0.656 & 0.188 & 0.438 \end{pmatrix}$$

步骤 9: calculate the evaluation vector $\hat{s}_i = (\hat{s}_{i1}, \hat{s}_{i2}, \cdots, \hat{s}_{in})$ which is closest to the evaluation vector of each round $s_i^t = (s_{i1}^t, s_{i2}^t, \cdots, s_{in}^t)$.

$\hat{s}_1 = (0.467, 0.268, 0.509, 0.467, 0.482)$

$$\hat{s}_2 = (0.391, 0.365, 0.365, 0.401, 0.648)$$

$$\hat{s}_3 = (0.380, 0.429, 0.379, 0.460, 0.562)$$

$$\hat{s}_4 = (0.570, 0.317, 0.412, 0.443, 0.458)$$

步骤 10：rank all the alternatives and select the best ones according to $\hat{s}_g$

$$\hat{s}_g = (0.453, 0.339, 0.421, 0.443, 0.536)$$

The order of five supply chain project $A_5 \succ A_1 \succ A_4 \succ A_3 \succ A_2$

**参考文献：**
[1] 张发明，郭亚军，张连怀.一种多阶段交互式群体评价方法[J].管理学报，2010,7（9）：1416-1420
[2] 张发明,郭亚军,易平涛 .序关系分析下的多阶段交互式群体评价方法[J].系统工程学报,2011,56(26):702-709
[3] 彭怡，胡杨.考虑群体一致性的动态群体决策方法[J].运筹与管理，2004,4（13）：69-72
[4] P.D. Liu, X. Zhang, Intuitionistic uncertain linguistic aggregation operators and their application to group decision making, Systems Engineering: Theory & Practice 32 (2012) 2704–2711.
[5] Z.S. Xu, Models for multiple attribute decision-making with intuitionistic fuzzy information, International Journal of Uncertainty, Fuzziness and Knowledge-Based Systems 15 (2007) 285–297.
[6] J.Q. Wang, Multi-criteria interval intuitionistic fuzzy decision-making approach with incomplete certain information, Control and Decision 21 (1263) (2006) 1253–1256.